\documentclass[amsmath, amssymb, aps, longbibliography, prl, reprint, superscriptaddress]{revtex4-2}
\usepackage{amsmath}
\usepackage{bibunits}
\usepackage{bm}
\usepackage[yyyymmdd]{datetime}
\usepackage{dcolumn}
\usepackage{graphicx}
\usepackage{physics}
\usepackage{siunitx}
\usepackage{hyperref}
\usepackage{upgreek}
\usepackage{xspace}
\hypersetup{ 
    colorlinks=true,
    linkcolor=blue,     
    urlcolor=blue,
    citecolor=blue,
}
\defaultbibliographystyle{apsrev4-2-titles}
\newcommand{\arctanh}[0]{\mathrm{arctanh}}

\begin{document}
\preprint{APS/123-QED}
\title{
A matter-wave Fabry-P{\'e}rot cavity in the ultrastrong driving regime
}
\author{Jeremy L. Tanlimco}
\author{Eber Nolasco-Martinez}
\thanks{Equal contribution.}
\author{Xiao Chai}
\thanks{Equal contribution.}
\author{S.~Nicole Halawani}
\thanks{Equal contribution.}
\author{Eric Zhu}
\thanks{Equal contribution.}
\affiliation{Department of Physics, University of California, Santa Barbara, California 93106, USA}
\author{Ivar Martin}
\affiliation{Materials Science Division, Argonne National Laboratory, Lemont, Illinois 60439, USA}
\author{David M. Weld}
\email{weld@physics.ucsb.edu}
\affiliation{Department of Physics, University of California, Santa Barbara, California 93106, USA}
\begin{abstract}
When the length of an optical cavity is modulated, theory predicts exponential concentration of energy around particular space-time trajectories. Viewed stroboscopically, photons in such a driven cavity propagate as if in a curved spacetime, with black hole and white hole event horizons corresponding to unstable and stable fixed points of the evolution. Such phenomena have resisted direct experimental realization due to the difficulty of relativistically accelerating massive cavity mirrors. We report results of an experiment which overcomes this limitation by exchanging the roles of light and matter. A matter wave endowed with quasi-relativistic dispersion is confined between two barriers made of light, one of which is periodically translated at speeds comparable to the matter wave group velocity. In this strongly-modulated cavity we observe the emergence of the predicted bright and dark fixed point trajectories, and demonstrate that changing the modulation waveform can vary the number of fixed points 
and exchange their stability character. We observe signatures of nontrivial dynamics beyond those predicted for photons, and attribute them to residual curvature in the dispersion relation. In addition to experimentally realizing and characterizing cavity dynamics in the ultra-strong driving regime, these results point the way to implementations of related dynamics in electro-optic materials, with potential applications in pulse generation and signal compression.
\end{abstract}
\maketitle
\begin{figure}[ht!]
    \centering
    \includegraphics[width=\linewidth]{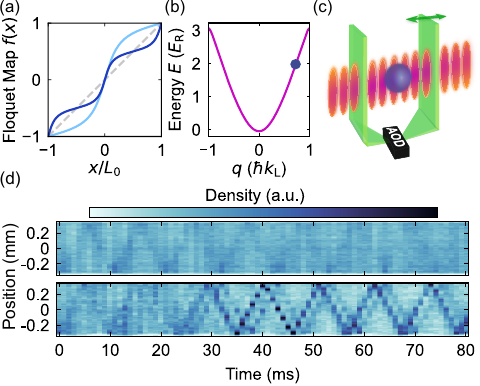}
    \caption{\textbf{Emergence of stable fixed point trajectory.} (a) The discrete Floquet map $f(x)$ of a cavity driven at the fundamental (light blue) and second-order (dark blue) resonances exhibits stable ($f'(x)<1$) and unstable ($f'(x)>1$) fixed points at the intersections with $f(x)=x$ (gray dashed). (b) Matter waves in the $D$ band of an optical lattice exhibit approximately linear (photonic) dispersion. (c) Lattice-trapped atoms between repulsive light sheets simulate a driven optical cavity. (d) Evolution of an initially diffuse atomic distribution in a static (top) and driven (bottom) cavity demonstrates drive-induced emergence of the stable fixed point trajectory.}
    \label{fig:emergence}
\end{figure}

\let\oldaddcontentsline\addcontentsline
\renewcommand{\addcontentsline}[3]{}
\textit{Introduction} | 
Two mirrors can create an optical cavity with a mode structure and frequency-dependent transmission that can be tuned by adjusting the geometry. This simple and ubiquitous optical instrument serves as a central element of lasers, spectrometers, clocks, and sensors~\cite{perot_application_1899,purcell_em_spontaneous_1946,ye_applications_2003}. 
Qualitatively new phenomena arise when a cavity's length is modulated at a frequency comparable to its free spectral range (the inverse of the round-trip travel time for photons).
In lasers, modulation can convert continuous radiation into intense pulses~\cite{henneberger_optical_1966,smith_phase_1967}.
In the regime of ultrastrong driving, even the electromagnetic vacuum state becomes unstable to entangled photon generation and exponential energy growth via the dynamical Casimir effect~\cite{moore_quantum_1970,law_resonance_1994,dalvit_renormalization-group_1998,dodonov_fifty_2020}. Furthermore, any initial field populating such a strongly-driven cavity is predicted to collect into short intense pulses, leaving the rest of the cavity in an ultra-cold state~\cite{martin_floquet_2019,wen_floquets_2022}. 
These phenomena share a convenient theoretical description based on the relationship between field configurations separated by a single period of modulation~\cite{martin_floquet_2019,wen_floquet_2018,koutserimpas_electromagnetic_2023}. Since vector potential solutions of the wave equation follow characteristics defined only by initial position and direction of propagation, a one-cycle Floquet map $f(x)$ captures the stroboscopic evolution for all positions $x$. The stable fixed points of this map (Fig.~\ref{fig:emergence}(a)) account for exponential energy concentration in temporally-compressed pulses. The stroboscopic evolution of the cavity field obeys the equations for wave propagation in a curved spacetime, with the stable and unstable fixed points corresponding to white hole and black hole event horizons, respectively~\cite{martin_floquet_2019}.

For energy concentration to overcome dissipation, however,
the mirror velocity must exceed $2c/Q$, where $c$ is the speed of light and $Q$ is the cavity mode quality factor~\cite{martin_floquet_2019}.
Furthermore, the mirror must repeatedly accelerate to this velocity from rest in a time of order $T=2L_0/c$ where $L_0$ is the average cavity length. 
For a lab-scale cavity, $T$ is in the nanosecond regime;
even for $Q\simeq10^{10}$, the required acceleration is on the order of a million times the earth's gravity~\footnote{In mode-locked lasers the operating conditions are not as demanding because the energy losses are compensated by externally provided population inversion.}. Thus, despite their fundamental interest and potential utility, cavities in this ultrastrong driving regime have resisted  experimental realization.

In this work, we experimentally address this challenge by exchanging the roles of light and matter. A quantum gas endowed by an optical lattice with a nearly relativistic dispersion relation~\cite{fujiwara_experimental_2018} and placed between two movable optical barriers (Figure~\ref{fig:emergence}(b-c)) realizes a matter-wave cavity in the ultrastrong driving regime, thanks not only to massless cavity mirrors but also to a reduction in the effective speed of light to less than one meter per second.
Using this approach we experimentally observe the predicted concentration of an initially diffuse atomic distribution to a single localized trajectory corresponding to the predicted stable fixed point. We show that a range of initial conditions converge to the same trajectory, and that the stroboscopic evolution agrees with theoretical predictions, with the two fixed points playing the roles of effective event horizons.
We demonstrate that increasing the drive frequency can increase the number of fixed point trajectories via higher-order resonances. Finally we show that a phase disruption in the driving waveform can reverse the stroboscopic time evolution and exchange the properties of the white-hole-like and black-hole-like fixed points.

\textit{Theory} | 
The emergence of fixed point trajectories can be intuitively understood by considering the simplest case of a resonantly driven cavity, where one mirror remains at rest at the origin and the other mirror moves on a path given by $L(t)=L_0(1+A\cos(\Omega t+\phi))$, where $\Omega=\pi c /L_0$. For this choice of $\Omega$, photons that encounter the mirror in its neutral position where $L(t)=L_0$ will return to that same point one cycle later. Two such fixed points exist, one stable and the other unstable.
If $\Omega$ is increased by an integer factor $p$, higher-order fixed points emerge, which concentrate energy in $p$ separate peaks. In contrast to the pulse formation in a mode-locked laser, these phenomena require no population inversion, cavity nonlinearity, or additional energy input beyond that required to move the mirror. 
Expanding a non-chiral cavity of length $L(t)$ into a chiral ring of circumference $2L(t)$ converts the second order wave equation to a first order one, whose initial conditions are fully specified by field values on $x\in[-L(0),L(0))$. 
After a single drive cycle, these fields are transported to new locations by the Floquet map $f(x):[-L,L)\mapsto[-L,L)$~\cite{martin_floquet_2019}. 
The complete evolution of the wavepacket is obtained by repeated iteration of this map. 
In the above example, for $A \ll 1$, the resonant Floquet map after $n$ cycles can be explicitly computed as
\begin{equation}
    \label{eqn:resonantFloquetMap}
    f^n(x)=\frac{L_0\phi}{\pi}+\frac{2L_0}{\pi}\arctan\left[e^{2\pi An}\tan\left(\frac{\pi x}{2L_0}-\frac{\phi}{2}\right)\right].
\end{equation}
Figure \ref{fig:emergence}(a) plots $f^1(x)$ for $p=1,2$, illustrating the stable and unstable fixed points.
The accumulation around the former and repulsion from the latter divides the set of initial positions $x$ into those whose stroboscopic velocity $f^{n}(x)-f^{n-1}(x)$ can exhibit both signs, or only one. At one boundary between these regions (the stable fixed point) the wavepacket can stroboscopically exit the ``one-way'' region but never enter, while at the other boundary (the unstable fixed point) it can only enter but never exit; the (un)stable fixed point is thus analogous to the event horizon of a (black) white hole~\cite{martin_floquet_2019}. More detailed discussion and derivation can be found in the supplemental information.

\textit{Experimental procedure} | The experiments begin by preparing a Bose-Einstein condensate (BEC) of $^7$Li with negligible thermal fraction in a crossed optical dipole trap and using a magnetic Feshbach resonance to tune contact interactions to zero. The atoms are adiabatically transferred to an optical lattice with spacing $d=532\,\mathrm{nm}$ and depth $V_\mathrm{L}=12.32\,E_\mathrm{R}$, where $E_\mathrm{R}=\hbar^2 k_\mathrm{L}^2/2M$ is the recoil energy of a lattice photon of wavevector $k_\mathrm{L}=\pi/d$.
To emulate a relativistic dispersion, atoms are transferred to the second excited Bloch band (\textit{D} band) of the lattice potential, an approach previously used to study relativistic harmonic motion~\cite{fujiwara_experimental_2018}. 
Band transfer is accomplished using a magnetic field gradient to initiate Bloch oscillations~
\cite{bloch_uber_1929,fujiwara_transport_2019,cao_transport_2020} while amplitude modulating the lattice with depth $1.58\,E_{\rm R}$ and frequency $\omega = 2\pi\times270\,\mathrm{kHz}$ to open an avoided crossing to the excited band~\cite{chai_continuously_2025}.
Upon ramping the force back to zero, the atomic wavepacket retains a residual $q_0=0.74\,\hbar k_\mathrm{L}$ quasimomentum, putting it near the most linear regime of the dispersion relation. The atoms are then confined between two repulsive optical barriers imposing an additional potential
\begin{equation}
    V_\mathrm{mirrors}(x,t) = V_\mathrm{M}\sum_{j=1}^2\exp\left[-\frac{\left(x-x_j(t)\right)^2}{2\sigma_\mathrm{M}^2}\right],
\end{equation}
with barrier height $V_\mathrm{M}=6.36\,\mathrm{E}_\mathrm{R}$ and width $\sigma_\mathrm{M}=10\,\upmu\mathrm{m}$ sufficient to suppress tunneling. The position of the second barrier is modulated with an acousto-optic deflector (AOD) to realize a matter-wave analogue of the resonantly driven cavity:
$
    x_2(t) = x_{2,0}+AL_0\cos(\Omega t-\phi),
$
for $AL_0$ the cavity modulation amplitude around the average position $x_{2,0}$ and $\Omega$ and $\phi$ the modulation frequency and phase, respectively. The resonant frequency $\Omega_0$ is set by the group velocity at $q_0$~\footnote{The numerically calculated Bloch band curvature changes sign at quasimomentum $q=0.8104\,\hbar k_\mathrm{L}$ due to band repulsion at the Brillouin zone center and edge. Above this inflection point, increasing quasimomentum counterintuitively \textit{decreases} group velocity.}:
\begin{equation}
    \Omega_0 = \frac{\pi v_\mathrm{g}}{L_0} = \frac{\pi}{L_0} \frac{\partial E(q)}{\partial q} \bigg|_{q=q_0}
\end{equation}
for $v_\mathrm{g}$ the group velocity, $E$ the Bloch band energy level, and $L_0=x_{2,0}-x_1=680.7\,\upmu\mathrm{m}$ the average cavity length. By fitting a triangle wave to the trajectory of the localized wavepacket in an undriven cavity, we determine $\Omega_0=95.1(5)\,\mathrm{Hz}$. 

\begin{figure}[ht!]
    \centering
    \includegraphics[width=\linewidth]{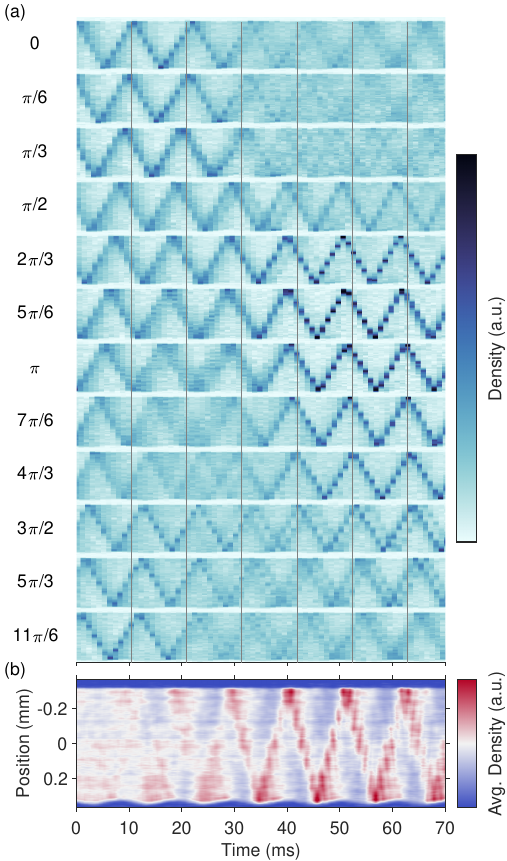}
    \caption{\textbf{Convergence of Various Initial Conditions to the Stable Fixed Point Trajectory.} (a) Absorption images of atomic density in the cavity over $80\,\mathrm{ms}$ for different modulation phases $\phi$, averaged over 3 repeats. The resulting trajectories are offset by $T/12$ to line up the boundary modulation and effectively scan initial position. Gray vertical lines indicate times at which the atoms' trajectory in a static cavity reflects from the upper wall, as extracted from a triangle fit. At early times, the offset trajectories are all out of phase with one another, but after several cycles of boundary modulation, they rephase. (b) The same temporally-offset absorption images overlaid on one another. An effectively uniform sampling of initial conditions coalesces onto the stable fixed point trajectory (red triangle wave) and is repelled from the unstable one (blue triangle wave) when subject to boundary modulation on the bottom wall.}
    \label{fig:phaseScan}
\end{figure}

\begin{figure*}[ht!]
    \centering
    \includegraphics[width=\linewidth]{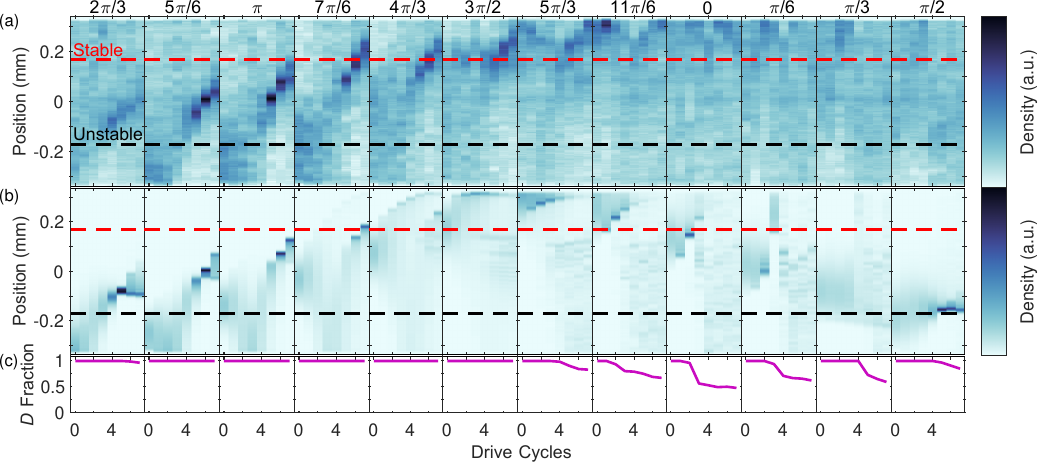}
    \caption{\textbf{Stroboscopic Motion toward an Event Horizon.} (a) Normalized atomic density imaged at the frequency of the boundary modulation for a range of initial conditions (modulation phases $\phi$) and averaged over 3 repeats.
    The red (black) dashed lines indicate the predicted (un)stable fixed points of the Floquet map. (b) Numerical simulation of a massive wavepacket initialized in the $2^\mathrm{nd}$ excited Bloch band subject to the same parameters as the experiment and to a Gaussian blur corresponding to the experimental imaging resolution.
    (c) Numerical simulation of $d$-band fraction. The boundary modulation imparts enough energy to the wavepacket to excite it into higher bands of the optical lattice, whose group velocities are no longer resonant with the drive (see Supplementary Information).
    }
    \label{fig:stroboscopic}
\end{figure*}

\textit{Results} | The first main result of this work is the experimental observation of the stable fixed point trajectory emerging from a diffuse initial state in a resonantly driven cavity. Figure~\ref{fig:emergence}(d) shows that while a uniformly filled cavity shows no special dynamics in the static case, if near-resonant ($\Omega=95.5\,\mathrm{Hz}$) modulation is applied to the same cavity with amplitude $A=1.45\times10^{-2}$, corresponding to a mirror velocity of more than a percent of the effective speed of light, 
we observe the emergence of a single concentrated trajectory in just a few drive cycles. This trajectory encounters the moving boundary when it is at the average length and moving inward, corresponding to the predicted stable fixed point of the one-cycle Floquet map. An additional fainter emergent trajectory not predicted by the photonic theory also emerges in this series of images and in some later data; we defer explanation of this interesting deviation from the predictions of the theory to discussion below~\cite{martin_floquet_2019}.

To probe the dependence of these dynamics on initial conditions, we prepare a more localized starting wavepacket at varying positions in the cavity. This is accomplished by adjusting the phase $\phi$ of the barrier modulation and offsetting trajectory measurements by an equivalent time delay. 
Results of such an experiment sampling $\phi$ in steps of $\pi/6$ across the full drive cycle and offsetting the resulting absorption images by $T/12$ (for $T=2\pi/\Omega$ the round-trip time) appear in Figure \ref{fig:phaseScan}(a). Across a range of starting points, the atomic trajectories depart from their different initial conditions and coalesce around a single trajectory, demonstrating the robustness of the emergent fixed point. Note that trajectories already close to the stable fixed point (modulation phase $0$) counter-intuitively experience stronger dissipation than those which start further away, and also that we again observe a faint additional emergent trajectory for a few initial conditions near $\phi=\pi/6$ and $5\pi/3$. To visualize these results in aggregate we average the phase-shifted atomic density distributions for all initial conditions and plot them using a bichromatic colormap in Figure \ref{fig:phaseScan}(b); the emergent blue areas indicate depletion of density around the unstable fixed point, and the emergent red areas indicates accumulation of density around the stable fixed point.

Imaging stroboscopically at the drive frequency, we observe that atomic wavepackets in a resonantly driven optical cavity reproduce the dynamics predicted by the one-cycle Floquet map. As shown in Figure \ref{fig:stroboscopic}(a), stroboscopically observed atoms from a representative sampling of initial positions drift away from the predicted unstable fixed point, indicated by the black dotted line, and toward the stable fixed point, indicated by the red one. Between the stable and unstable fixed points, the stroboscopic velocity assumes only one sign, in agreement with the curved spacetime interpretation of the dynamics and the identification of fixed points with black and white hole event horizons. Results of numerical integration of the time-dependent Schrodinger equation, shown in Fig. \ref{fig:stroboscopic}(b), quantitatively reproduce the observed approach to the stable fixed point for most initial conditions. For drive phases near 0 we observe smearing of the atomic wavepacket and the disappearance of sharp trajectories. The numerical model confirms analytic expectations of the physical reason for this behavior: repeated encounters with an approaching mirror blue-shift the wavepacket to higher quasimomenta, leading at some point to an encounter with the edge of the Brillouin zone and loss from the \textit{D} band via Bragg scattering. Figure \ref{fig:stroboscopic}(c) supports this interpretation, showing the evolution of the numerically calculated \textit{D}-band fraction for each phase. 

\begin{figure}[b!]
    \centering
    \includegraphics[width=\linewidth]{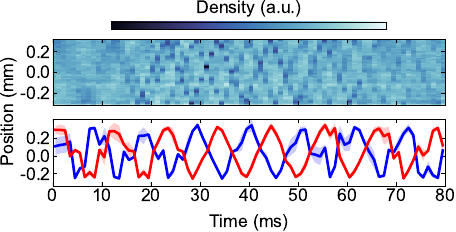}
    \caption{\textbf{Emergence of Multiple Stable Fixed Point Trajectories from Higher-Order Cavity Resonances.} (Top) Modulating the boundary of the cavity at twice the fundamental frequency results in two stable fixed point trajectories at late times. (Bottom) Fitting the density distributions to two doubly-reflected Gaussian distributions recovers the expected $\pi$ phase difference between the two trajectories. Shaded region is one standard deviation width of the atomic cloud.}
    \label{fig:2ndOrder}
\end{figure}

Due to the massless cavity mirrors and the extremely low effective speed of light, this matter-wave cavity can easily reach even stronger driving regimes, achieving higher-order resonances and creating pulse trains of multiple localized wavepackets. To probe the simplest such case, we drive an initially homogeneously filled cavity at twice its fundamental resonance $\Omega=191\,\mathrm{Hz}$ and half the previous modulation amplitude $A=7.27\times10^{-3}$ to preserve the maximum velocity of the wall $A\Omega$. Under these conditions we observe the emergence of two stable fixed point trajectories, shown in Figure \ref{fig:2ndOrder}. 

\begin{figure}[ht!]
    \centering
    \includegraphics[width=\linewidth]{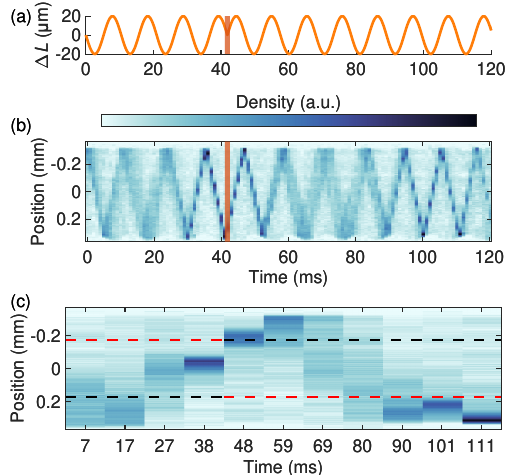}
    \caption{\textbf{Stroboscopic Time Reversal.} (a) The modulation waveform $\Delta L=AL_0\cos(\Omega t-\phi(t))$ with a sharp phase step of $\pi$ after four complete drive cycles and a modulation depth $A=2.9\times10^{-2}$. (b) An atomic wavepacket initialized at the unstable fixed point initially evolves toward the stable one, but once the stable and unstable fixed points switch roles, it returns to its original trajectory. (c) Sampling stroboscopically at approximately the drive period, the wavepacket moves from one fixed point to the other and back again.}
    \label{fig:loschmidt}
\end{figure}

Since this platform is not limited by the mechanical motion of a physical mirror, a variety of boundary modulation waveforms beyond simple sinusoidal driving can be explored, including even phase-discontinuous waveforms. In Figure \ref{fig:loschmidt}, we demonstrate a simple application of this flexibility by implementing an abrupt change of the modulation phase. Such a phase jump is expected to shift the action of the Floquet map, reversing the direction of flow for a range of positions. Measurements of the resulting dynamics under this waveform show that an atomic wavepacket close to the unstable fixed point trajectory initially approaches the stable one; after a sudden $\pi$ phase shift in the boundary modulation switches the stable and unstable fixed points, the wavepacket reverses course and returns to its original trajectory. This demonstration of stroboscopic time reversal points the way to possible schemes for time-domain signal compression and decompression; an arbitrary initial condition localized into short pulses by the boundary modulation can be revived in the same number of cycles by phase-shifting the modulation~\cite{martin_floquet_2019}.

\textit{Discussion} | Taken together, these results clearly demonstrate that the driven matter-wave cavity realizes the exotic dynamics predicted for a modulated optical cavity in the regime of ultrastrong driving, including the exponential emergence of single and multiple fixed-point trajectories and their detailed dependence on modulation waveforms. 
The data also reveal phenomena beyond the predictions of the simplest driven-cavity model. All of these can be understood to arise due to deviations from perfectly linear dispersion. We have already discussed the way in which such deviations near the Brillouin zone edge can smear wavepackets near the stable fixed point, as shown in Fig.~3. A more subtle effect is the observation, mentioned but not explained above, of additional emergent stable trajectories for some modulation phases. These arise from the fact that any curvature in the dispersion relation means that reflection from a moving barrier changes the group velocity, something which would also occur in a real optical cavity filled with any medium other than vacuum. As a result, the evolution of the driven cavity depends on the velocity of the modulated wall as well as its position. This explains the observation of additional fixed point trajectories at the nodes of wall velocity (the extrema of the position modulation, corresponding to phases $\pi/2$ and $3\pi/2$) as well as the dynamical instability of the stable fixed point trajectory (which for resonant modulation aligns with the point of largest momentum transfer from the moving barrier). Numerical simulations quantitatively confirm this explanation of the effects of imperfectly linear dispersion (Supplemental Information).

The dynamics we observe, along with the possibility of their realization in other (\textit{e.g.} electro-optic) platforms, suggest new avenues for optical signal compression and multiplexing, energy concentration, and cavity cooling~\cite{martin_floquet_2019,wen_floquets_2022,sinatkas_electro-optic_2021,galiffi_photonics_2022}. The experiments we describe can also be extended in a variety of intriguing directions. The rich variety of dynamics afforded by nonlinear dispersion can be further augmented by Floquet engineering~\cite{fujiwara_transport_2019,cao_transport_2020,chai_continuously_2025,huang_nondispersing_2021,su_dynamics_2024}, 
providing an infinite set of tunable dispersion relations exhibiting sign changes in group velocity, nontrivial topology, and even asymmetry across the Brillouin zone~\cite{minguzzi_topological_2022}.  Likewise, the massless cavity mirrors can be driven with a diverse range of modulation waveforms, of which the abrupt phase jump demonstrated above is only one example. Combining these two generalizations would enable the study of dynamical phenomena like Fermi acceleration, recently observed in a similar context~\cite{barontini_observation_2025}. Feshbach tuning of contact interactions would enable simulation of many-body states. The insertion into the cavity of a partially transmitting mirror, acting as a coherent beamsplitter or a Josephson junction~\cite{bernhart_observation_2025}, would enable experiments probing the dynamics of coupled driven cavities.
If the atoms can be initialized in the cavity's absolute ground state, driving to higher Fock states could realize the dynamical Casimir effect~\cite{lamoreaux_demonstration_1997,law_resonance_1994,dalvit_renormalization-group_1998,wilson_observation_2011,jaskula_acoustic_2012,nation_colloquium_2012}, wherein pairwise excitations spontaneously arise from vacuum. These and related experiments could potentially be used to probe phenomena such as black-hole information dynamics~\cite{hawking_breakdown_1976}, Unruh radiation~\cite{unruh_notes_1976,fulling_nonuniqueness_1973}, and quantum chaos~\cite{lapierre_floquet_2025}.




\section{Acknowledgments}\label{sec:rollcredits}
We acknowledge Xuanwei Liang and Alec Cao for experimental assistance, Aashish Clerk for helpful discussions, and Yifei Bai for a critical reading of the manuscript. We acknowledge support from the Army Research Office (W911NF-22-1-0098 and W911NF-23-1-0291), the Noyce Foundation, the Eddleman Quantum Institute, and the NSF QLCI program through Grant No. OMA-2016245. E.N.-M. acknowledges support from the UCSB NSF Quantum Foundry through the Q-AMASEi program (Grant No. DMR-1906325). S.N.H. acknowledges support from the NSF NRT program under grant 2152201. I.M. was supported by the Materials Science and Engineering Division, Basic Energy Sciences, Office of Science, US Department of Energy.


\bibliography{modulatedCavity.bib}


\clearpage
\let\addcontentsline\oldaddcontentsline

\widetext
\begin{center}
    \textbf{\large Supplementary Information: \\A matter-wave Fabry-P{\'e}rot cavity in the ultrastrong driving regime}
\end{center}
\setcounter{figure}{0}
\renewcommand{\figurename}{Fig.}
\makeatletter 
\renewcommand{\thefigure}{S\@arabic\c@figure}
\renewcommand{\theHfigure}{S\@arabic\c@figure}
\makeatother
\renewcommand{\thesection}{\arabic{section}}
\renewcommand{\thesubsection}{\thesection.\arabic{subsection}}
\renewcommand{\thesubsubsection}{\thesubsection.\arabic{subsubsection}}
\setcounter{equation}{0}
\renewcommand{\theequation}{{S}\arabic{equation}}

\makeatletter
\setcounter{section}{0} 
\renewcommand{\thesection}{\arabic{section}} 
\makeatother

\makeatletter
\renewcommand\section{%
\@startsection
{section}%
{1}%
{\z@}%
{0.8cm \@plus1ex \@minus .2ex}%
{0.5cm}%
{%
\normalfont\small\bfseries
\centering
}%
}%
\def\@seccntformat#1{\csname the#1\endcsname.\quad}
\makeatother

\tableofcontents

\clearpage
\section{Methods}\label{sisec:methods}
\begin{figure}[ht!]
    \centering
    \includegraphics[width=0.6\textwidth]{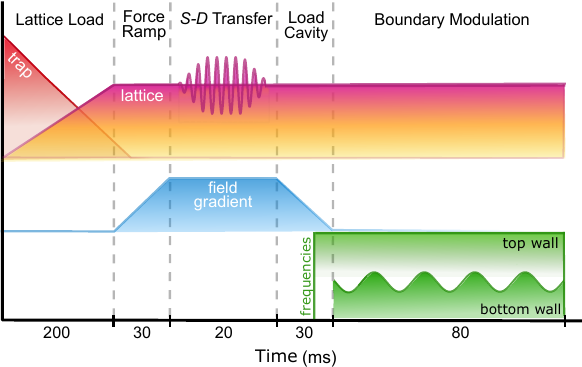}
    \caption{\textbf{Experimental Sequence.} Bose-condensed atoms are adiabatically loaded into an optical lattice over 200ms as the crossed dipole trap ramps down. Resonant amplitude modulation of the lattice along with an applied field gradient transfers the atoms from the \textit{S} band to the \textit{D} band, endowing them with quasi-relativistic dispersion. Blue-detuned light sheets generated by an acousto-optic deflector (AOD) confine the matter wave as the force ramps to zero. Sinusoidal modulation of the AOD frequencies generates the desired boundary modulation of the bottom wall with phase $\phi$. Full details of the experimental sequence can be found in Methods.}
    \label{fig:sequence}
\end{figure}
We begin our experiment with a Bose-Einstein condensate (BEC) of $2\times10^5$ $^7$Li atoms prepared in the $|F,m_F\rangle=|1,1\rangle$ state and in the ground motional state of a $\lambda=1064\,\mathrm{nm}$ crossed optical dipole trap. After evaporative cooling to negligible thermal fraction at a higher field, the uniform vertical magnetic field is reduced to $543.6\,\mathrm{G}$ to set the s-wave scattering length to zero~\cite{pollack_extreme_2009}. Atoms are then adiabatically transferred over $200\,\mathrm{ms}$ from the crossed dipole trap to a $1064\,\mathrm{nm}$ horizontal optical lattice with $110\,\upmu\mathrm{m}$ $4\sigma$ diameter and a depth of $12.32\,E_\mathrm{R}$, populating a momentum distribution centered at quasimomentum 0 of the lowest lattice band (denoted \textit{S} following orbital notation). Over $25\,\mathrm{ms}$, we ramp up a magnetic field gradient along the same axis to effect a tilted lattice potential and initiate Bloch oscillations ~\cite{bloch_uber_1929,dahan_bloch_1996} with period $T_\mathrm{B}=10.70\,\mathrm{ms}$. We then apply a $270\,\mathrm{kHz}$ amplitude modulation of the optical lattice intensity in a $4\,\mathrm{ms}$ trapezoidal pulse ($0.1\,\mathrm{ms}$ rise and fall), which creates a Landau-Zener avoided crossing \cite{zener_non-adiabatic_1932} between the ground band and the second excited (\textit{D}) band to transfer the atoms to the latter band~\cite{denschlag_bose-einstein_2002}. After an additional $8.6\,\mathrm{ms}$ of Bloch oscillation in the \textit{D} band, we ramp the magnetic field gradient back to 0 over another $25\,\mathrm{ms}$, leaving the final momentum at $2.74\,\hbar k_\mathrm{L}$, where $k_\mathrm{L}=2\pi/\lambda$ is the lattice wave vector. During this ramp, we also turn on a $532\,\mathrm{nm}$ light sheet $1.2\,\mathrm{mm}$ away from the atoms so that they reflect coherently from it about $1\,\mathrm{ms}$ after the field gradient ramp has finished. $4\,\mathrm{ms}$ after that, a second $532\,\mathrm{nm}$ light sheet turns on $0.6\,\mathrm{mm}$ away from the original position of the atoms, confining the moving atomic wavepacket between the two Gaussian potential barriers of height $6.36\,E_\mathrm{R}$ and $4\sigma$ axial diameter of $40\,\upmu\mathrm{m}$. The probability of tunneling through these barriers is negligible, and their position can be modulated with an acousto-optic deflector. After a variable hold time in the static cavity potential we can perform absorption imaging of the atomic density to map out a triangle-wave trajectory of atoms bouncing back and forth between the barriers, from which we can extract the resonant frequency $\Omega_0$ of the static cavity. To achieve a more uniform initial filling of the cavity motional states (as in Figures 1 and 4a), we hold the BEC in the lattice $S$ band for $100\,\mathrm{ms}$ before ramping up the force and in the static cavity potential for $68\,\mathrm{ms}$ before beginning boundary modulation. 

\section{Modulated Cavity Model}\label{sisec:model}

In this section we describe the mathematical structure of the dynamics which occur in a modulated optical cavity. The main text then presents the mapping between such a modulated optical cavity and the quasi-relativistic matter-wave cavity which is realized and characterized experimentally in this work.

\subsection{Conformal Map of the Dynamical Cavity}\label{sec:conformal}
Consider an optical cavity bounded by ideal mirrors, one of which is stationary at $x=0$ and the other of which is evolving with trajectory $z(t)$. We ignore the transverse coordinates $\bm{\hat{\mathrm{y}}},\bm{\hat{\mathrm{z}}}$ and assume that the waves propagate along only the $\bm{\hat{\mathrm{x}}}$ axis. If the electric field is polarized along $\bm{\hat{\mathrm{z}}}$ (without loss of generality) and the scalar potential is identically zero in the absence of charges, then
\begin{equation}
    \label{eqn:cavityFields}
    \begin{split}
        \mathbf{A}&=\mathcal{A}\bm{\hat{\mathrm{z}}}\\
        \mathbf{E}&=-\frac{\partial\mathcal{A}}{\partial t}\bm{\hat{\mathrm{z}}}\\
        \mathbf{B}&=-\frac{\partial\mathcal{A}}{\partial x}\bm{\hat{\mathrm{y}}}.
    \end{split}
\end{equation}
The magnitude of the magnetic vector potential $\mathcal{A}$ within the cavity evolves (in the Coulomb gauge) according to the wave equation
\begin{equation}
    \label{eqn:staticCavityWaveEquation}
    \frac{\partial^2\mathcal{A}}{\partial t^2}-\frac{\partial^2\mathcal{A}}{\partial x^2}=0.
\end{equation}
Pioneering theoretical work~\cite{moore_quantum_1970,martin_floquet_2019} showed that the problem of a cavity with a dynamically evolving boundary can be mapped conformally to the static cavity thus:
\begin{equation}
    \label{eqn:conformalMap}
    \begin{split}
        t+x&=F(s+w)\\
        t-x&=F(s-w).
    \end{split}
\end{equation}
The wave equation is conveniently automorphic under the action of this conformal map: 
\begin{equation}
    \label{eqn:waveEquationAutomorphism}
    \begin{split}
        \frac{\partial^2\mathcal{A}}{\partial s^2}-\frac{\partial^2\mathcal{A}}{\partial w^2}=0.
    \end{split}
\end{equation}
Moreover, $F$ can be chosen to map the dynamical and static cavity boundaries to one another: $(t,0)\to(s,0)$ and $(t,z(t))\to(s,1)$, from which follows
\begin{equation}
    \label{eqn:swBoundaryConditions}
    \begin{split}
        t&=F(s)\\
        t+z(t)&=F(s+1)\\
        t-z(t)&=F(s-1),
    \end{split}
\end{equation}
and, defining the inverse function $R=F^{-1}$,
\begin{equation}
    \label{eqn:RBootstrap}
    R(t+z(t))-R(t-z(t))=2.
\end{equation}
Eq. (\ref{eqn:RBootstrap}) serves as a convenient bootstrap for mapping solutions of the static cavity to those of the dynamical one and for propagating forwards or backwards in time from a known solution interval, as long as the mirror motion is subluminal. The wave equation Eq. (\ref{eqn:waveEquationAutomorphism}) in $s$ and $w$ results in solutions of the form
\begin{equation}
    \label{eqn:staticCavityWaveSolutions}
    \mathcal{A}(s,w)=\mathcal{A}_+(s+w)+\mathcal{A}_-(s-w),
\end{equation}
that is, linear combinations of left-($-x$) and right-($+x$) moving wavepackets of arbitrary functional form, respectively. Requiring the (now static) boundary conditions $\mathcal{A}(s,0)=\mathcal{A}(s,1)=0$ requires that for all $s$, $\mathcal{A}_+(s)=-\mathcal{A}_-(s)$ and $\mathcal{A}_+(s)=\mathcal{A}_+(s+2)$; invoking Eq. (\ref{eqn:RBootstrap}) and these boundary conditions in the original problem gives rise to
\begin{equation}
    \label{eqn:dynamicCavityWaveSolutions}
    \tilde{\mathcal{A}}(t,x)=\mathcal{A}_+(R(t-x))-\mathcal{A}_+(R(t+x)).
\end{equation}
As expected, this is a linear combination of functions of $t\pm x$, so it solves the original wave equation Eq. (\ref{eqn:staticCavityWaveEquation}); the periodic boundary conditions of $\mathcal{A}_+$ and Eq. (\ref{eqn:RBootstrap}) ensure that $\mathcal{A}(t,0)=\mathcal{A}(t,z(t))=0$ as required. This Dirichlet boundary condition requires that the incoming and reflected waveforms are equal and opposite; moreover, since this property is invariant under the action of a Lorentz boost, it applies both to static and moving walls, as long as the latter does not move superluminally ($|\dot{z}|<1$). 

\subsection{Chiral Solution on a Ring}\label{sisec:ring}
The above method, while viable, requires careful accounting of the left- and right-moving waveforms but can be further simplified by unfolding the cavity bounded by $[0,z(t)]$ into a ring $[-z(t),z(t))$ wherein right-moving waves in $[0,z(t))$ map to themselves in $[0,z(t))$ and left-moving waves in $[0,z(t)]$ map to right-moving waves in $[-z(t),0]$. When right-moving waves in the ring encounter the ``boundary'' at $z(t)$, they change sign and re-enter at $-z(t)$ to close the loop; in other words, the waveform $\mathcal{A}$ only moves in the $+x$ direction along the loop with ``boundary conditions'' requiring discontinuous sign changes at $0$ and $\pm z(t)$. Under this transformation, Eq. (\ref{eqn:staticCavityWaveEquation}) simplifies to a first-order chiral equation
\begin{equation}
    \label{eqn:chiralWaveEquation}
    \frac{\partial\mathcal{A}}{\partial t}-\frac{\partial\mathcal{A}}{\partial x}=0
\end{equation}
for $x\in[-z(t),z(t)]$, which can be easily solved with the method of characteristics. The resulting characteristics are the null lines along which the vector potential $\mathcal{A}$ is constant, with the exception of the discontinuous sign changes at $0$ and $\pm z(t)$. Moreover, since the differential equation is first-order, the initial-value problem only requires the vector potential $\mathcal{A}(0,x)=\mathcal{A}_0(x)$ and not its derivatives; in this sense, the dynamics are Markovian, and the vector potential at some later time $t$ can be determined by tracing the right-moving null lines $t-x=\mathrm{const.}$ back to $t=0$.

Under the coordinate transformation to the chiral ring domain, the solutions Eq. (\ref{eqn:dynamicCavityWaveSolutions}) reduce to
\begin{equation}
    \label{eqn:chiralCavityWaveSolution}
    \tilde{\mathcal{A}}(t,x)=\mathcal{A}_+(R(t-x))\mathrm{sgn}(x).
\end{equation}
Already we can qualitatively deduce the compressing character of a trajectory $z(t)$ that changes direction, periodic or not: cavity expansion ($\dot{z}>0$) adds empty segments to the ring along which $\mathcal{A}=0$, while cavity compression ($\dot{z}<0$) interferes $\mathcal{A}$ with its sign-changed tail on the other side of the moving boundary $z(t)$. The long-time limit of periodic driving, then, should accumulate $\mathcal{A}$ around increasingly narrow wavepackets~\cite{martin_floquet_2019}.

\subsection{Periodic Driving}\label{sisec:periodic}
The above analysis applies to a generic (and subluminal) boundary trajectory $z(t)$ by packaging the details of that trajectory into the conformal mapping Eq. (\ref{eqn:conformalMap}). In the special case of periodic driving, the field dynamics reduce to a considerably simplified picture characterized by a discrete Floquet map and stroboscopic observation with the same period as the drive; the fixed points of the Floquet map (if they exist) identify trajectories along which the field accumulates in the long-time limit.

\subsubsection{Floquet Map and Its Fixed Points}\label{sisec:FloquetMap}
As previously stated, the evolution of the vector potential inside the cavity reduces to a problem of tracing null lines back to the initial state. For periodic driving, the cavity by definition evolves the same way in each drive cycle and returns to its initial state after every drive cycle; as a result, we only require knowledge of how null lines evolve within a given drive cycle (of period $T$) and a Floquet map function mapping the position of a null line at time $t_0$ to its position in the next drive cycle $t_0+T$:
\begin{equation}
    \label{eqn:FloquetMap}
    f(x):x_{t_0}\to x_{t_0+T}
\end{equation}
A multi-period map can be defined $f^{(p)}(x):x_{t_0}\to x_{t_0+pT}$, as well as an inverse $g=f^{-1}$ that propagates the null line backward in time (\textit{i.e.}, discretizing the bootstrapping action of $R$ as given in Eq. (\ref{eqn:RBootstrap})).

A discrete map can have stroboscopic fixed points defined by $x_0=f(x_0)$ (or for higher-order fixed points characterized by cycles of length $p$, $x_0=f^{(p)}(x_0)$); such fixed points can be stable ($|f'(x_0)|<1$) or unstable ($|f'(x_0)|>1$), depending on whether small deviations $\delta$ compound or decay. Fixed points of order $p$ necessarily imply at least $p$ distinct fixed points: if $x_0$ is a fixed point, then so is $f(x_0)$, and so on. The fixed points actually represent trajectories, insofar as the map is discrete and the observation consequently stroboscopic; the fixed point trajectories simply trace the same null lines in each cycle.

For instance, consider a cavity driven at the lowest resonance:
\begin{equation}
    \label{eqn:fundamentalResonantDriving}
    z(t)=L_0\left(1+A\sin(\Omega_0 t)\right)
\end{equation}
where $2L_0=cT=2\pi c/\Omega_0$ indicates the distance traversed by the null line during one cycle. During this cycle, the boundary is at its neutral position $L_0$ twice; once with $|\dot{z}|>0$ and once with $|\dot{z}|<0$. Null lines that encounter the boundary at this position at time $t=nT$ ($t=(n+1/2)T$ in the second case) will experience negative (positive) Doppler shift that serves to expand (contract) the waveform. Negative (positive) Doppler shifts in turn result in unstable (stable) fixed points. In the long-time limit, trajectories close to the unstable fixed point are repelled, while those close to the stable one accumulate~\cite{martin_floquet_2019}.

The existence of fixed points persists for slightly off-resonant drives as long as the trajectory $z(t)$ contains two points at $L_0$, that is, $\mathrm{min}[z(t)]<L_0<\mathrm{max}[z(t)]$. Here, $L_0$ may differ from the average cavity length $L$, as long as $A$ is large enough to admit two points at $L_0$, which then occur closer together or further apart.

Typically, the Floquet map $f:x(t_0)\to x(t_0+T)$ (abbreviated $x_0\to x_1$ for convenience) cannot be determined explicitly. For instance, for a moving boundary close to the fundamental resonance given by Eq. (\ref{eqn:fundamentalResonantDriving}), the null line will encounter the boundary once in a drive cycle at time $t_\mathrm{m}$ given by
\begin{equation}
    \label{eqn:tmImplicit}
    t_\mathrm{m}=
    \begin{cases}
        t_0+\frac{z\left(t_\mathrm{m}\right)-x_0}{c} & z'\left(t_\mathrm{m}\right)>0\\
        t_0+T-\frac{z\left(t_\mathrm{m}\right)+x_1}{c} & z'\left(t_\mathrm{m}\right)<0
    \end{cases}
\end{equation}
from which can be derived the the length of the trajectory
\begin{equation}
    \label{eqn:L0Implicit}
    2L_0=x_1-x_0+2z\left(t_\mathrm{m}\right).
\end{equation}
The above equations only define $x_1$ and $t_\mathrm{m}$ implicitly. However, if $z\left(t_\mathrm{m}\right)$ were known explicitly (with no reference to $t_\mathrm{m}$), the Floquet map and its inverse would arise directly from Eq. (\ref{eqn:L0Implicit}):
\begin{equation}
    \label{eqn:fundamentalFloquetMap}
    \begin{split}
        x_1&\equiv f\left(x_0\right)=x_0-2\left[z\left(t_\mathrm{m}\right)-L_0\right]\\
        x_0&\equiv g\left(x_1\right)=x_1+2\left[z\left(t_\mathrm{m}\right)-L_0\right].
    \end{split}
\end{equation}

\subsubsection{Perturbative Solution of the Resonant Drive}\label{sisec:weakDrive}
\begin{figure}[ht!]
    \centering
    \includegraphics{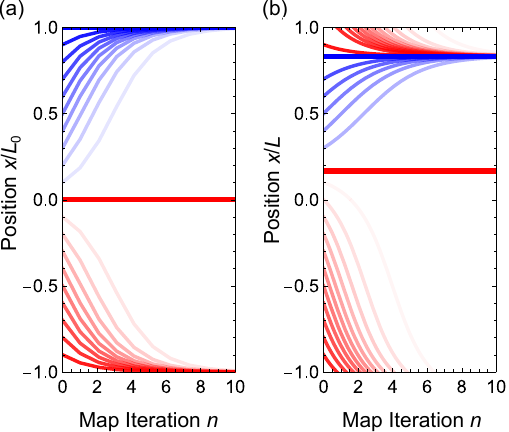}
    \caption{\textbf{Perturbative solutions for the Floquet maps.} (a) The resonant case (Eq. (\ref{eqn:resonantMultiPeriodMap}) for $A=L_0/10$ and $\phi=0$. The unstable fixed point lies at $x_\mathrm{unstable}=0$ (thick red), while other initial points drift to the stable fixed point at $x_\mathrm{stable}=\pm L_0$ (thick blue). Right- and left-moving stroboscopic trajectories are indicated with blue and red, respectively, at varying opacity. (b) The off-resonant case (Eq. (\ref{eqn:offResonantMultiPeriodMap})) for $A=L/10$ and $\Delta=L/20$. The unstable fixed point shifts to $x_\mathrm{unstable}=L/6$ and the stable one to $x_\mathrm{stable}=5L/6$.}
    \label{fig:perturbativeSolutions}
\end{figure}
For a weak drive near the fundamental resonance (\textit{i.e.}, $A\ll 1$ and $L\approx L_0$ for $L$ the average cavity length), the self-consistent Eq. (\ref{eqn:tmImplicit}) can be approximated $t_\mathrm{m}=t_0+\left(L_0-x_0\right)/c$, giving the Floquet map
\begin{equation}
    \label{eqn:fundamentalFloquetMapPerturbative}
    \begin{split}
        f\left(x_0\right)&=x_0+2L_0-2z\left(t_0+\frac{L_0-x_0}{c}\right)\\
        &=x_0+2\left(L_0-L\right)-2AL\sin\left(\frac{\pi\left(L_0+ct_0-x_0\right)}{L}\right)
    \end{split}
\end{equation}
Since the cavity is driven close to resonance, the action of the Floquet map is small, so the iteration step $n$ can be approximated as a continuous variable.
\begin{equation}
\label{eqn:fundamentalFloquetDerivative}
    f\left(x_0\right)-x_0\approx\frac{\mathrm{d}x}{\mathrm{d}n}=2\left(L_0-L\right)-2AL\sin\left(\frac{\pi\left(L_0+ct_0-x\right)}{L}\right)
\end{equation}
Additionally, we can identify the cavity detuning $\Delta=\frac{L-L_0}{L}$ and denote the reference time
\begin{equation}
    \label{eqn:referenceTime}
    t_0=\frac{1}{c}\left(\frac{L\phi}{\pi}-L_0\right),
\end{equation}
relating the reference time to the phase of the boundary modulation $\phi$ relative to the trajectory. This greatly simplifies Eq. (\ref{eqn:fundamentalFloquetDerivative}):
\begin{equation}
    \label{eqn:fundamentalFloquetDerivativeSimplified}
    \frac{\mathrm{d}x}{\mathrm{d}n}=2L\left[-\Delta+A\sin\left(\frac{\pi x}{L}-\phi\right)\right].
\end{equation}
At resonance, the stable fixed point occurs at $x_\mathrm{stable}=L_0(1+\phi/\pi)$ and the unstable one at $x_\mathrm{unstable}=L_0\phi/\pi$; near resonance, they will occur at
\begin{equation}
    \label{eqn:detunedFixedPointConditions}
    \begin{split}
        x_\mathrm{stable}&=L+\frac{L}{\pi}(\phi-\theta)\\
        x_\mathrm{unstable}&=\frac{L}{\pi}(\phi+\theta),
    \end{split}
\end{equation}
for $\theta=\arcsin\left(\frac{\Delta}{A}\right)$. The explicit form of the multi-period map can be determined by integration of Eq. (\ref{eqn:fundamentalFloquetDerivative}):
\begin{equation}
    \label{eqn:fundamentalFloquetIntegral}
    \begin{split}
        n&=\int_{x_0}^{x_n}\frac{\mathrm{d}x}{2L\left[-\Delta+A\sin\left(\frac{\pi x}{L}-\phi\right)\right]}\\
        &=-\frac{1}{\pi\sqrt{A^2-\Delta^2}}\left[\arctanh\left[\frac{A-\Delta\tan\left(\frac{\pi x}{2L}-\frac{\phi}{2}\right)}{\sqrt{A^2-\Delta^2}}\right]\right]_{x_0}^{x_n}
    \end{split}
\end{equation}
Rearrangement gives the multi-period map
\begin{equation}
    \label{eqn:offResonantMultiPeriodMap}
    x_n^{(\Delta\ne0)}=\frac{L\phi}{\pi}+\frac{2L}{\pi}\arctan\left[\frac{A+\sqrt{A^2-\Delta^2}\tanh\left[\frac{\sqrt{A^2-\Delta^2}n\pi}{L}+\arctanh\left[\frac{\Delta\tan\left(\frac{\pi x_0}{2L}-\frac{\phi}{2}\right)-A}{\sqrt{A^2-\Delta^2}}\right]\right]}{\Delta}\right]
\end{equation}
Substitution of $\theta=\arcsin\left(\frac{\Delta}{A}\right)$ recovers the fixed points (Eq. \ref{eqn:detunedFixedPointConditions}):
\begin{equation*}
    \sin\left(\frac{\pi x_n^{(\Delta\ne0)}}{2L}-\frac{\phi+\theta}{2}\right)\cos\left(\frac{\pi x_0}{2L}-\frac{\phi-\theta}{2}\right)=e^{2\pi An\cos\theta}\sin\left(\frac{\pi x_0}{2L}-\frac{\phi+\theta}{2}\right)\cos\left(\frac{\pi x_n^{(\Delta\ne0)}}{2L}-\frac{\phi-\theta}{2}\right)
\end{equation*}
The limit $\theta\to0$ or re-evaluation of the integral in the case of exact resonance $\Delta=0$ gives
\begin{equation}
    \label{eqn:resonantMultiPeriodMap}
    x_n^{(\Delta=0)}=\frac{L_0\phi}{\pi}+\frac{2L_0}{\pi}\arctan\left[e^{2\pi An}\tan\left(\frac{\pi x_0}{2L_0}-\frac{\phi}{2}\right)\right].
\end{equation}
Equivalently,
\begin{equation*}
    \sin\left(\frac{\pi x_n^{(\Delta=0)}}{2L_0}-\frac{\phi}{2}\right)\cos\left(\frac{\pi x_0}{2L_0}-\frac{\phi}{2}\right)=e^{2\pi An}\sin\left(\frac{\pi x_0}{2L_0}-\frac{\phi}{2}\right)\cos\left(\frac{\pi x_n^{(\Delta=0)}}{2L_0}-\frac{\phi}{2}\right).
\end{equation*}
As expected, $x_\mathrm{unstable}=L_0\phi/\pi$ and $x_\mathrm{stable}=L_0(1+\phi/\pi)$ (modulo $2L_0$) are fixed points, while $x_n^{(\Delta=0)}\to L_0(1+\phi/\pi)$ for all other $x_0$ in the $n\to\infty$ limit. The off-resonant case has the same qualitative behavior for slightly shifted fixed points.

This perturbative analysis confirms the qualitative observation that stable and unstable fixed points arise in pairs, and that nonzero detuning of the cavity with respect to the fundamental resonance (or its overtones) merely shifts the location of the fixed points as long as the driving amplitude $A$ exceeds the detuning $\Delta$.

\subsection{General Relativistic Interpretation}\label{sisec:relativity}
\begin{figure}[ht!]
    \centering
    \includegraphics{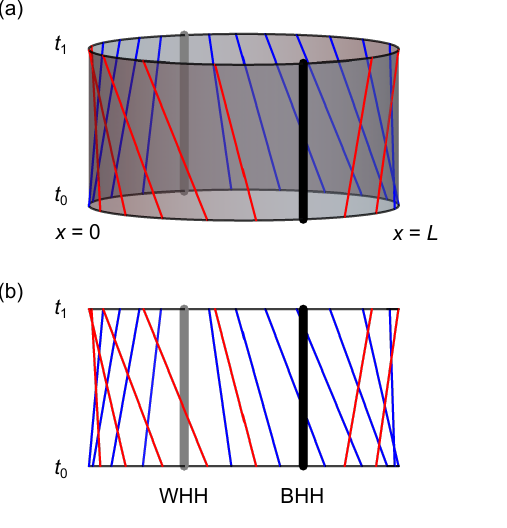}
    \caption{\textbf{Stroboscopic relativity.} (a) Representation of the one-step resonant Floquet map (Eq. (\ref{eqn:resonantMultiPeriodMap})) with an additional phase shift in the chiral ring geometry. Black and gray mark the unstable and stable fixed points, respectively; blue and red indicate left- and right-movers in the original cavity geometry. (b) Projection of the Floquet map back onto the original nonchiral cavity, with the same color scheme as in (a). The stroboscopic ``null lines'' can point in either direction outside the central region; in between the two fixed points, stroboscopic velocity has only one sign. This is analogous to black- and white-hole event horizons (corresponding to the unstable and stable fixed points, respectively), across which light can travel in only one direction.}
    \label{fig:stroboscopicRelativity}
\end{figure}
The stroboscopic ``velocity'' $\mathrm{d}x/\mathrm{d}n$ of the Floquet map can be interpreted as originating from a nontrivial spacetime curvature. Not to be confused with the actual group velocity $c$ of the vector potential waveform traversing the null lines, this ``velocity'' changes sign at the fixed points of the Floquet map and varies continuously between them. Projecting the ring geometry back onto the one-dimensional cavity, 
there exist 
regions where the stroboscopic velocity can only have one sign (see Fig.~\ref{fig:stroboscopicRelativity}). This has a direct analogy to the behavior of light near black hole event horizons; light entering the black hole can continue through the event horizon, but light attempting to exit from behind the event horizon will approach 0 stroboscopic velocity; likewise, the other fixed point exhibits the opposite behavior | that of a white hole, from which light may exit but never enter~\cite{martin_floquet_2019}.
The perturbative solution Eq. (\ref{eqn:fundamentalFloquetDerivativeSimplified}) to a simple sinusoidal drive does not match any canonical general relativistic model, even disregarding the lack of a singularity; however, the character of the event horizon can certainly be tuned with a different choice of modulation $z(t)$.


\section{Numerical Simulation}\label{sisec:simulation}
The numerical simulations presented in Figure 3(b) of the main text are performed by integrating the one-dimensional time-dependent Schr\"odinger equation (TDSE),
\begin{equation}
    \mathrm{i} \hbar \pdv{t} \Psi(x,t)=\left[- \frac{\hbar^2}{2 M} \pdv[2]{x} + V_{\rm lattice}(x,t) + V_{\rm mirrors}(x,t)\right]\Psi(x,t),
\end{equation}
where $M$ is the atomic mass, $\Psi(x,t)$ is the single-particle wavefunction, and the potentials are
\begin{equation}
    V_\mathrm{lattice}(x,t) = - V_\mathrm{L}\left(1+\alpha\cos(\omega t)\right)\cos\left(k_\mathrm{L}x\right)
\end{equation}
\begin{equation}
    V_\mathrm{mirrors}(x,t) = V_\mathrm{M}\sum_{j=1}^2\exp\left[-\frac{\left(x-x_j(t)\right)^2}{2\sigma_\mathrm{M}^2}\right].
\end{equation}
The initial condition is a superposition of Bloch waves in the $D$ band with a full-width-half-maximum (FWHM) quasimomentum spread of $0.1\,\hbar k_{\rm L}$. Since the initial wave packet is not Heisenberg-limited, we include a free-space expansion phase for each momentum component to match the experimental wavefunction size of $\SI{106}{\micro\meter}$ (FWHM) in real space. To precisely determine the center of the initial quasimomentum distribution, we perform TDSE simulation with different $q_0$ for a static cavity using measured parameters of the barriers. When $q_0=0.736\,\hbar k_{\rm L}$, the simulated cavity frequency $\Omega_0$ matches the experimental value of $\SI{95.5}{\hertz}$. We then use this value as the initial quasimomentum center to perform simulations for the modulated cavity. The initial real-space center of the wavefunction is the center of the cavity. The wall modulation is initiated at different delays to match the experimental conditions. To faithfully sample the wavefunction under modulations, we use grid sizes of $\SI{0.1}{\micro\second}$ in time and $\SI{5.7}{\nano\meter}$ in space.

\begin{figure}
    \centering
    \includegraphics[width=\linewidth]{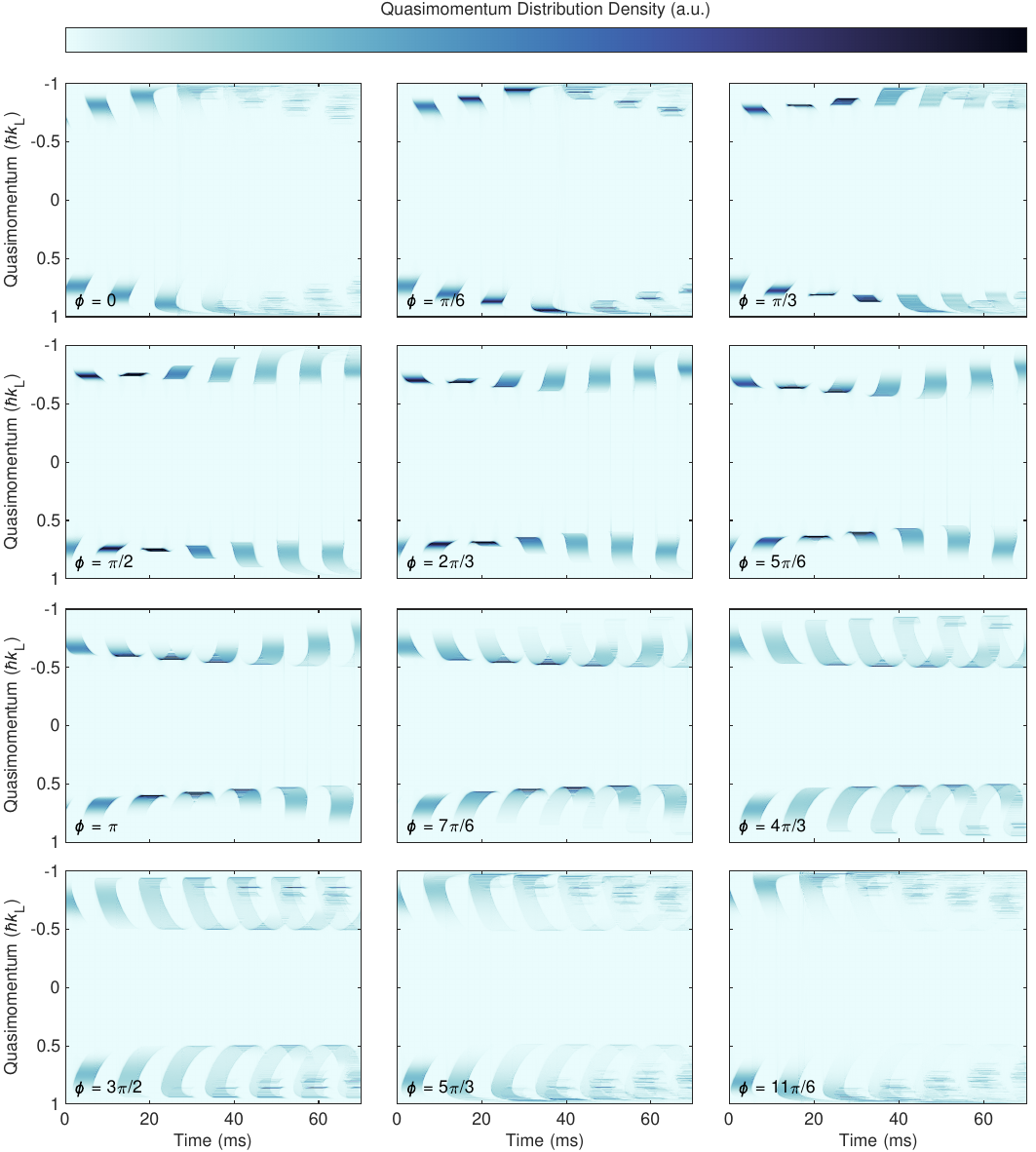}
    \caption{\textbf{$D$ band Quasimomentum evolution.} In addition to the numerical simulation of real-space dynamics shown in Fig.~3 in the main text, here we present the time evolution of quasimomentum distribution in $D$ band under the same simulation parameters. For some modulation phases, the modulations kick the atoms towards the Brillouin zone edge, which leads to Landau-Zener tunneling to higher bands.}
    \label{fig:qt}
\end{figure}

\begin{figure}
    \centering
    \includegraphics[width=\linewidth]{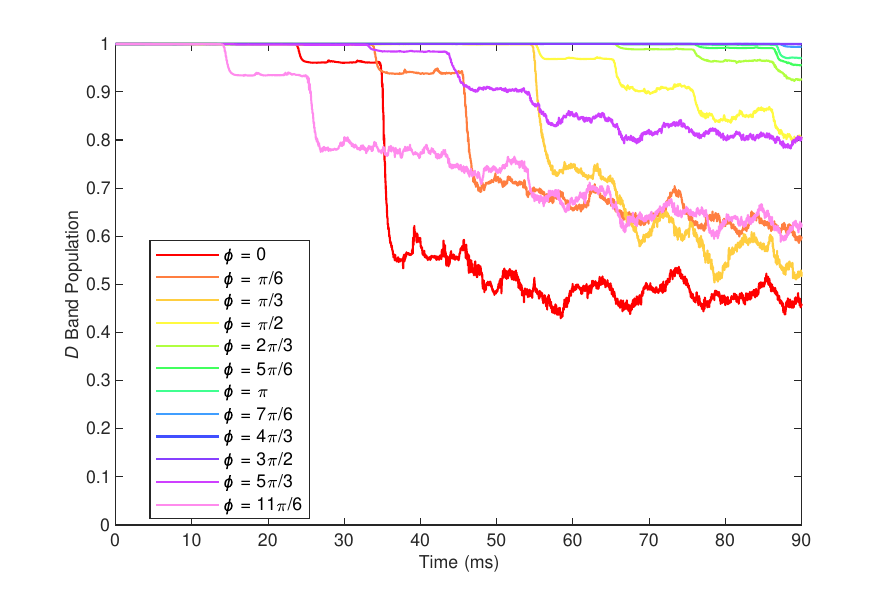}
    \caption{\textbf{$D$ Band fraction.} The numerical simulation of the full evolution of $D$ band fraction as a supplement to Fig.~3(c) in the main text.}
    \label{fig:bandt}
\end{figure}

Besides the real-space stroboscopic dynamics shown in Fig.~3(b), the simulation also reveals that the quasimomentum explores a significant fraction of the Brillouin zone, as shown in Figure ~\ref{fig:qt}, supporting our hypothesis that significant changes in the group velocity impart velocity dependence to an originally only position-dependent Floquet map. The immediate dissipation of wavepackets following trajectories close to the stable fixed point ($\phi=0$) correlates with the quasimomentum reaching the Brillouin zone edge and consequent population transfer to the higher ($F$) band (Fig.~\ref{fig:bandt}). Since the stable fixed point corresponds to the trajectory that encounters the boundary with maximum blue-shifting, it follows that its momentum should increase with each drive cycle. Meanwhile, the appearance of spurious fixed points (phases $\phi=3\pi/2,5\pi/3$) in Figures \ref{fig:phaseScan} and \ref{fig:stroboscopic} correlates with a quasimomentum distribution spanning half the Brillouin zone. This occurs for trajectories that encounter the wall near the point of its closest approach and where the instantaneous momentum gain is negligible but whose resulting fixed point is locally stable to deviations in momentum.

\clearpage
\end{document}